\documentclass[aps, pra, twocolumn, notitlepage,superscriptaddress,nofootinbib   
]{revtex4-1}
\usepackage{amssymb,latexsym,amsmath}
\usepackage{bm}
\usepackage[normalem]{ulem}
\usepackage{amssymb,amsmath}
\usepackage{graphicx,xcolor}
\usepackage[titletoc, title]{appendix}
\usepackage[caption=false,lofdepth,lotdepth]{subfig}
\usepackage{hyperref}   
\hypersetup{%
	pdfpagemode=FullScreen,
	pdfstartpage=1,
	pdfmenubar=true,
	pdftoolbar=true,
	colorlinks = true,
	linkcolor=blue,
	citecolor=blue,
	urlcolor=blue,
	bookmarksopen=false
}

\newcommand{\rev}[1]{{\color{black}#1}}

\usepackage{soul}

\begin{document}

\title{Dynamics of a massive superfluid vortex in \texorpdfstring{$r^k$}{rk} confining potentials}
\author{Andrea Richaud}
\affiliation{Scuola Internazionale Superiore di Studi Avanzati (SISSA), Via Bonomea 265, I-34136, Trieste, Italy}
\author {Pietro Massignan}
\email{pietro.massignan@upc.edu}
\affiliation{Departament de F\'isica, Universitat Polit\`ecnica de Catalunya, Campus Nord B4-B5, E-08034 Barcelona, Spain}
\author{Vittorio Penna}
\affiliation{Dipartimento di Scienza Applicata e Tecnologia, Politecnico di Torino, Corso Duca degli Abruzzi 24, I--10129 Torino, Italy}
\author {Alexander L.\ Fetter}
\email{fetter@stanford.edu}
\affiliation {Departments of Physics and Applied Physics, Stanford University, Stanford, CA 94305-4045, USA}

\date{\today}

\begin{abstract}

\rev{
We study the motion of a superfluid vortex in condensates having different background density profiles, ranging from parabolic to uniform. 
The resulting effective point-vortex model for a generic power-law potential $\propto r^k$ 
can be experimentally realized 
with recent advances in optical-trapping techniques. Our analysis encompasses both  empty-core and filled-core vortices. In the latter case, the}
vortex acquires a mass due to the presence of 
distinguishable atoms located in its core.
The axisymmetry allows us to reduce the coupled dynamical equations of motion  to a single radial equation with an effective potential $V_{\rm eff}$.  In many cases, $V_{\rm eff}$ has a single minimum, where the vortex precesses uniformly.
\rev{The dynamics of the vortex and the localized massive core arises from the dependence  of the energy on the radial position of the vortex and from the $r^k$ trap potential.  We find that a}
 positive vortex with small mass orbits in the positive direction, but the sense of precession can reverse as the core mass increases.  
 \rev{Early experiments and theoretical studies on two-component vortices found some  qualitatively similar behavior.}

\end{abstract}

\maketitle

\section{Introduction}

Superfluid vortices have been of great interest ever since Feynman's seminal  article in 1955 \cite{Feynman1955}.  The creation of ultracold atomic Bose-Einstein condensates (BECs) in 1995 \cite{Pethick, Pitaevskii}  broadened the original focus on liquid $^4$He to include many new possibilities.  
The first BEC vortex was in a two-component condensate with two trapped hyperfine states of $^{87}$Rb \cite{Matthews1999,Anderson2000}, although most subsequent experiments \rev{ \cite{Madison2000,Raman2001,Leanhardt2002,Scherer2007,Kwon2016}} studied simpler one-component BECs.  

In these mixtures, each component had its own resonant frequency and could be imaged separately, allowing nondestructive visualization of the large filled core, whose radius was larger than the
\rev{optical resolution of the imaging system.}
Consequently, it was feasible to study the precession of two-component  vortices in real time \cite{Anderson2000}.  In contrast, the empty core of  a one-component vortex typically has a radius smaller than the wavelength of the imaging light and is observable only after free expansion by turning off the trap. Various methods subsequently allowed direct real-time observation of precession of  a one-component vortex, the most direct visualizing the dynamics through expansion of successive small fractions of the condensate \cite{Freilich2010,Serafini2017}. Collisions between vortices have also been studied in great detail \cite{kwon2021sound,Richaud2022_Collisions}.
Soon after the first experiments, theoretical studies used a time-dependent variational Lagrangian to study the precession of  one-component and two-component vortices \cite{Lundh2000, McGeeHolland}, although little detailed comparison was made with the  experiments.  

 Various theoretical works have studied the dynamics of massive vortices  over the last few years \cite{Richaud2020,PRA_Griffin,Richaud2021,Ruban2022,Doran2022}. In these systems, the vortex in component $a$ surrounds  a localized massive core in component $b$, assuming interaction constants that would strongly favor phase separation of the two components in a uniform system.
Our previous works focused on motion in a two-dimensional flat trap with a circular boundary \cite{Richaud2020,Richaud2021}.  Here we extend our model to include a trap with a power-law potential $r^k$.  For simplicity, we study a single vortex in a Thomas-Fermi condensate $a$.  This model allows us to interpolate between the usual harmonic trap with $k=2$ and the flat trap \cite{Navon2021,Dalibard_DMD_2021}  in the limit $k\to \infty$. 

The precession of an off-center vortex around the axis of an harmonic trap 
requires a non-trivial theoretical description (see Ref.~\cite{Greoszek2018} and references therein) due to the 
spatially-varying
condensate's density profile. 
\rev{Several works over the past twenty years considered models which included different features, like image vortices and core sizes that depend on the local density~\cite{Jackson1999,Lundh2000,Svidzinsky2000PRL,Svidzinsky2000PRA,KimFetter,McGeeHolland,Anglin2002,Sheehy2004,Jezek2008,Ednilson2016,Esposito2017,Biasi2017}.
Our model incorporates both features. For small $k$ (especially the harmonic trap with $k = 2$), we find that the vortex precession rate can decrease and even reverse direction as the localized core mass increases. It is notable that earlier experimental and theoretical studies~\cite{Anderson2000,McGeeHolland} each found evidence of such reversal of precession, even though these studies were near the onset of bulk phase separation.


In our previous study of a single vortex in a flat potential with a circular boundary~(see Sec.~II.A of Ref.~\cite{Richaud2021}),  the Lagrangian for a vortex with a massive localized core had a term linear in the vortex velocity along with the usual kinetic energy that is quadratic in the vortex velocity. This linear term is familiar from the Lagrangian of a massive point particle in an external electromagnetic field.  For this system in a flat trap,  there is an effective uniform magnetic field $\bm B = -2\pi n_a\hbar \hat{\bm z}$, where $n_a$ is the two-dimensional number density of the background component. Our present analysis  includes a nonuniform $r^k$ trapping potential, and the effective magnetic field also becomes nonuniform. 
More importantly, the corresponding effective vector potential $\bm A$ now appears in the Hamiltonian as a synthetic gauge field that depends explicitly on the choice of trap potential. 
This formulation generalizes the familiar Hamiltonian for massless vortices~(see, for example Sec.~157 of Ref.~\cite{Lamb}).}


The paper is structured as follows:
 Section~\ref{sec:massless} relies on the Thomas-Fermi model in a power-law trap to find the condensate density for  a single-component BEC.   This result allows us to obtain a  time-dependent variational Lagrangian, based on a trial function describing a single quantized vortex in a power-law potential, along  with its opposite-sign image outside the condensate. The Lagrangian characterizes the dynamics of a single  vortex, which here yields uniform circular precession. 
 Section~\ref{sec:massive} adds the massive localized core to obtain the Lagrangian for a massive point vortex.  In addition to the usual kinetic energy of the core mass, it also has a term linear in the vortex velocity.  We discuss the analogy with the electromagnetic Lagrangian for a charged particle and find the associated synthetic vector potential and synthetic magnetic field. 
 The dynamics of a single massive point vortex typically involves uniform circular precession along with small stable oscillations around the local minimum in the effective potential.  In some cases, however, this minimum disappears, and the vortex moves  to the outer boundary. 
 Positive massless vortices precess in the positive direction around the trap center, but we find that as their mass increases the precession frequency can reverse  sign. 
   We end with conclusions and outlook in Sec.~\ref{sec:conclusions}.

\section{Thomas-Fermi  model for single-component BEC}
\label{sec:massless}
\rev{A single-component BEC is described, at the mean-field level, by the familiar Gross-Pitaevskii (GP) model
\begin{equation}
\label{eq:GPE_1_condensate}
i\hbar\frac{\partial \Psi_a}{\partial t} =\! \!\left(\!-\frac{\hbar^2\nabla^2}{2m_a} + V_{\rm tr} 
+ g_{aa}|\Psi_a|^2\!\right)\Psi_a, 
\end{equation}
where $g_{aa}=2\sqrt{2\pi}\hbar^2 a_{aa}/(m_{a} d_z)$ represents the effective interaction in the quasi-2D system, $a_{aa}$ being the component-$a$ s-wave scattering length, $m_{a}$ its atomic mass, and $d_z$ the harmonic-oscillator length along the $z$-direction \cite{Hadzibabic2011}. Our analysis thus focuses on an effective 2D system (lying on the $xy$ plane), as the possible degrees of freedom along the $z$-axis are assumed to be frozen due to the strong confinement along that direction.

In the strongly-interacting Thomas-Fermi (TF) regime}, for any axisymmetric potential $V_{\rm tr}(r)$, the TF condensate density $n_a(r) $  satisfies
\begin{equation}\label{TF}
    \mu_a = V_{\rm tr}(r) + g_{aa}n_a(r).
    \end{equation}
Here $\mu_a = g_{aa}n_0$ is the chemical potential of the $a$ component, and $n_0=n_a(0)$ is the density at the center of the trap.   
If $n_a(r)$ vanishes at the TF radius $R$, then Eq.~(\ref{TF}) implies that $g_{aa}n_0 = V_{\rm tr}(R)$. For a  power-law potential $\propto r^k$ with $k>1$, the trap potential can be rewritten as 
\begin{equation}\label{Vtr}
V_{\rm tr}(r) = g_{aa}n_0 (r/R)^k.
\end{equation}
By construction, the TF density  $n_a(r) = n_0[1-(r/R)^k]$ vanishes at the TF radius. The total number of particles is $N_a = \int d^2 r \,n_a(r)$, with the resulting $k$-dependent central density 
\begin{equation}\label{na}
n_0 = \frac{k+2}{k}
\frac{N_a}{\pi  R^2},
\end{equation}
where the  numerical factor $(k+2)/k $ varies smoothly  in going from a harmonic trap  ($k = 2 $) to a flat trap ($k\to \infty$). 

\subsection{Time-dependent variational Lagrangian}

To study the dynamics of a vortex in a power-law trap, we rely on the time-dependent variational Lagrangian, which has proved valuable in many aspects of BEC physics \cite{PerezGarcia1996},   instead of the more complete  GP equation \rev{(\ref{eq:GPE_1_condensate})}. In this approach, one takes a trial wave function $\Psi_a(\bm r,\bm \rho)$ for the $a$ component  that depends on the vortex position $\bm \rho$ as a time-dependent parameter, where we use $\bm r =(r,\theta) $ as a general coordinate and $\bm \rho = (\rho,\phi)$ for the position of the vortex.  
Use the trial function $\Psi_a$
to evaluate the Lagrangian $L_a=T_a-E_a$ for the $a$ component,  
where
\begin{equation}
T_a[\Psi_a] =\frac{i\hbar}{2}\int d^2 r\left(\Psi_a^*\frac{\partial\Psi_a}{\partial t} - \frac{\partial \Psi_a^*}{\partial t }\Psi_a\right)
\end{equation}
and 
\begin{equation}\label{E}
 E_a[\Psi_a] = \int d^2r\left(\frac{\hbar^2}{2m_a}|\bm \nabla \Psi_a|^2 +V_{\rm tr}|\Psi_a|^2 + \frac{g_{aa}}{2}|\Psi_a|^4\right)
\end{equation}
depend on the coordinate of the vortex  $\bm \rho$ through $\Psi_a$.
 
We use the TF model, with   $\sqrt{n_a(r)}$ as the amplitude of the trial function $\Psi_a$. We assume a single vortex at $\bm \rho$ with dimensionless charge $q=\pm 1$ and an opposite-charge image vortex at $\bm \rho'= \bm{\hat\rho}R^2/\rho$ outside the condensate.  This image is necessary for a flat trap and it facilitates the comparison for general values of $k$.
Let $S(\bm r,\bm\rho)$ be the angle between the vector $\bm r-\bm \rho$ and the $\hat{\bm x}$ axis. We choose the trial function  
\begin{equation}\label{Psia}
    \Psi_a(\bm r) = \sqrt{n_a(r)}\,e^{iq[S(\bm r,\bm \rho) -  S(\bm r,\bm\rho')]}
\end{equation}
which includes the phase of the vortex and its image.
Unless otherwise specified, we will assume that $q=+1$ in the following. 
\rev{We remark that the assumed density profile $n_a(r)$ does not include spatial density variations arising from the presence of the vortex. This is justified because vortex cores in atomic BECs  have a radius of the order of the component-$a$ healing length $\xi_a=\hbar/\sqrt{2m_ag_{aa}n_0}$,  and the latter is generally much smaller than the TF radius $R$ \cite{Pethick,Pitaevskii}.}

The evaluation of $T_a$ and $E_a$ follows as in Ref.~\cite{KimFetter}.  With our trial function (\ref{Psia}), we find 
\begin{equation}\label{T}
  T_a=\hbar q \,\dot{\bm \rho}\times \hat{\bm z}\cdot\int d^2 r\,n_a(r)\frac{\bm r - \bm \rho}{|\bm r - \bm \rho|^2}  
\end{equation}
plus a similar term for the image vortex at $\bm\rho'$.
The integral in (\ref{T}) is a vector that must lie along $\hat{\bm\rho}$ by symmetry, so that  
\begin{equation*}
  T_a=\hbar q \,\dot{\bm \rho}\times \hat{\bm z}\cdot \hat{\bm \rho} \int d^2 r\,n_a(r)\frac{r \cos\theta' - \rho}{r^2  - 2r\rho\cos\theta' +\rho^2},  
\end{equation*}
where $\theta' = \theta-\phi$.
The  angular integral gives $-(2\pi/\rho)\Theta(\rho - r)$ where $\Theta$ is a unit step function that vanishes for $r > \rho$.  A straightforward calculation gives 
\begin{equation}\label{Ta}
        T_a( \rho,\dot{\phi}) = -q\hbar  \pi n_0 R^2\,\dot\phi\,\tau(\tilde \rho)
\end{equation}
where $\tilde \rho = \rho/R$ is the dimensionless scaled radial vortex  position and 
\begin{equation}
    \tau(\rho) = 2\int_0^\rho rdr\,(1-r^k) = \rho^2-\frac{2 \rho^{k+2}}{k+2}
\end{equation}
 is a dimensionless function of $ \rho$. We now drop the tilde and treat $\rho$ as dimensionless.  By construction, note that
 $0\leq\tau(\rho)<1$ and \begin{equation}\label{tau'}
 \frac{\tau'(\rho)}{\rho} = \frac{2n_a(\rho)}{n_0}  =2(1-\rho^k).
 \end{equation}
 A similar analysis for the contribution of the image vortex gives $\dot\phi$ multiplied by a constant because the image vortex lies outside the condensate.  Since this contribution is a perfect time derivative, we can ignore it and retain only Eq.~(\ref{Ta}).

The remaining term is the incremental energy of the vortex in Eq.~(\ref{E}).  With our TF trial function, $E_a$ is the  kinetic-energy density of the vortex and its image integrated over the condensate density
\rev{
\begin{equation}\label{Evor}
E_a = \frac{1}{2}\int d^2 r \,m_a n_a(r) \left|\bm v(\bm r-\bm \rho) - \bm v(\bm r-\bm \rho')\right|^2,    \end{equation}
where 
\begin{equation}
\bm v(\bm r-\bm \rho) = \hat{\bm z} \times \frac{\bm r-\bm \rho}{|\bm r -\bm \rho|^2}
=\hat{\bm z}\times \bm \nabla \ln|\bm r-\bm \rho|
\end{equation}
is  the dimensionless flow velocity of a vortex at $\bm \rho$, and $\bm v(\bm r - \bm \rho')$ is the corresponding flow velocity of the image vortex at $\bm \rho'$.
Here, the last form uses the alternative representation involving the stream function $\chi(\bm r-\bm \rho) =\ln|\bm r - \bm \rho|$.

The stream function gives the total flow velocity as $\bm v(\bm r) =(\hbar/m_a)\,\hat{\bm z}\times \left[\bm \nabla \chi (\bm r-\bm \rho) -\bm \nabla\chi(\bm r - \bm \rho')\right]$.
We follow~\cite{KimFetter} and find}
\begin{equation}\label{Ea}
E_a(\rho) = \frac{\hbar^2 \pi n_0}{m_a}\epsilon( \rho),
\end{equation}
where 
\begin{eqnarray}
\epsilon(\rho) & = & \ln\left(\frac{1-\rho^2}{\delta(\rho)}\right)  +\frac{\rho^k}{2}\ln\left(\frac{\delta(\rho)^2}{2\rho^2}\right)\nonumber \\
&+& \frac{\rho^k}{2}\left\{ H_{k/2} + H_{-k-2} +(-1)^k\left[\hat B_{-1}  -\hat B_{-1/\rho}\right] \right. \nonumber \\ 
&+& \left.B_{\rho}(-k-1,0) +\frac{2\rho^2}{k+2}\left[2\hat F(\rho^2) -\rho^2\hat F(\rho^4)\right] \right\}\nonumber\\
\nonumber \\
&+&\frac{1}{2\rho^k}\left\{\hat B_{\rho^2}
- \hat B_{\rho}+ (-1)^k\left[\hat B_{-\rho}-\hat B_{-\rho^2}\right]\right\}
\label{eq:epsilon_of_rho}
\end{eqnarray}
with $\rho$ again dimensionless.  Here $H_n $ denotes the harmonic number, $B_\rho(k,\alpha)$ is the incomplete beta function [and we often used the shorthand notation $\hat B_{\rho}\equiv B_\rho(k+2,0)$],
$$\hat F(\rho) = {}_2F_1\left(1,\textstyle\frac{k}{2}+1,\frac{k}{2}+\frac{1}{2};\rho\right)$$
is a hypergeometric function. 

\rev{Here we use a density-dependent  cutoff at the vortex core $\delta(\rho) =\delta_0\sqrt{n_0/n_a(\rho)}$ with  $\delta_0$ a constant of order $\xi_0/R\approx 0.01$. Such a cutoff ensures the convergence of the integral (\ref{Evor}), as it effectively excludes a neighborhood of radius $\delta$ centered at $\bm{\rho}$ where the background TF density $n_a$ is finite and the quantity $\left|\bm v(\bm r-\bm \rho) - \bm v(\bm r-\bm \rho')\right|^2$ features a non-integrable singularity \cite{KimFetter}. In the spirit of Ref.~\cite{McGeeHolland}, the specific functional dependence assumed for $\delta(\rho)$ mimics that of the condensate healing length $\xi$. This dependence  incorporates the radial dependence of the vortex characteristic width into the variational model and has a nontrivial impact on the ensuing massless  vortex dynamics.}

\subsection{Dynamical motion of a massless vortex}
\begin{figure}[h!]
    \centering
    \includegraphics[width=0.8\linewidth]{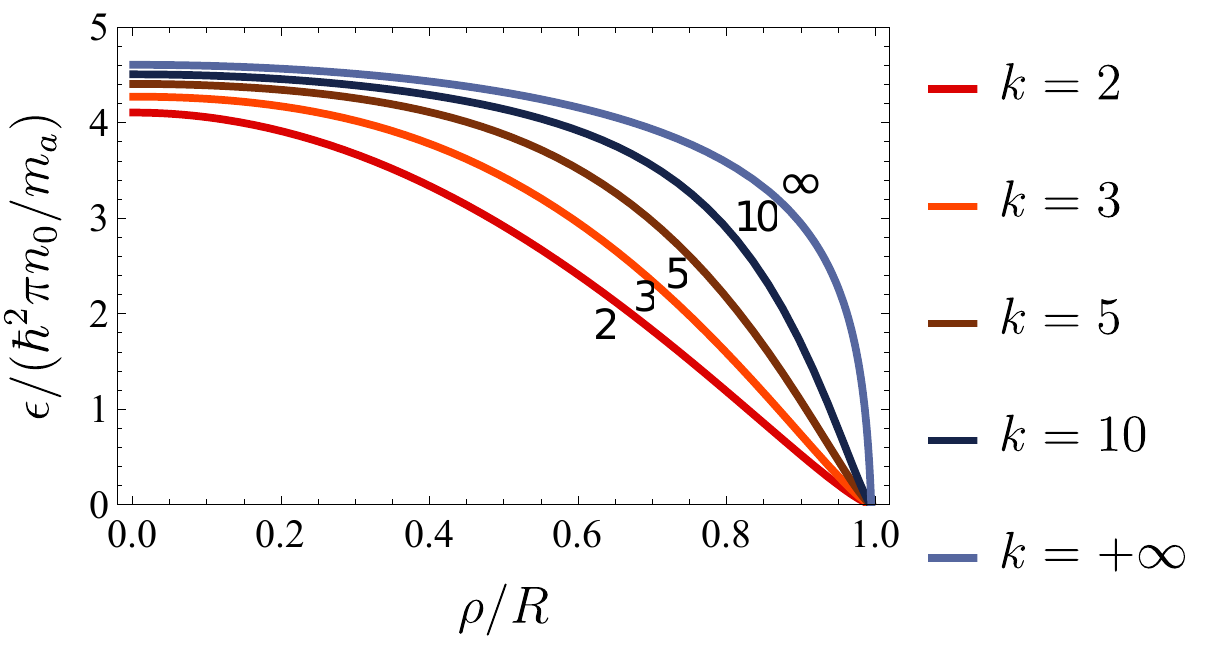}
    \caption{
    Dimensionless function $\epsilon(\rho)$ defined in Eq.~(\ref{eq:epsilon_of_rho}) for different values of $k$ (given by the numbers next to the lines).
    }
    \label{fig:Epsilon_vs_rho}
\end{figure}

It is now convenient to introduce dimensionless variables, with $R$ as the length scale, $m_a R^2/\hbar$ as the time scale and $\hbar^2 \pi n_0/m_a$ as the energy scale.  In this way we have the  very simple dimensionless Lagrangian for the pure $a$-component vortex
\begin{equation}\label{La}
    L_a(\rho,\dot\phi) = -q\tau(\rho)\dot\phi -\epsilon(\rho).
\end{equation}
This Lagrangian conserves the angular momentum  $l = \partial L_a/\partial \dot\phi= -q\tau(\rho)$, so that the vortex precesses at fixed $\rho$.  
Note that by construction a massless positive vortex has always $l<0$.
The precession rate $ \dot \phi$ follows from $\partial L_a/\partial \rho = -q\tau'(\rho)\dot\phi -\epsilon'(\rho) = 0$
 with the dimensionless angular speed
\begin{equation}\label{Omegaa}
    \Omega = \dot\phi = -\frac{\epsilon'(\rho)}{{2q\rho(1-\rho^k)}}.
\end{equation}
The quantity $\epsilon'(\rho)$ is always negative \rev{(see Fig.~\ref{fig:Epsilon_vs_rho})} while $\rho(1-\rho^k)$ is positive. As such, a massless  vortex precesses in the same sense as its circulation $q$.

For a flat potential ($k\to\infty$), we have  $\epsilon_\infty = \ln(1-\rho^2)$, reproducing the usual result~\cite{Richaud2020,Richaud2021}
\begin{equation}
\label{eq:Omega_infty}
\Omega_\infty =\frac{\hbar\, q}{m_a R^2}\,\frac{1}{1-\rho^2},
\end{equation}
now rewritten in conventional units.  For the harmonic potential with $k = 2$, we have  
\begin{equation}\label{eps2}
 \epsilon_2 = (1-\rho^2)\left[\frac{(1+\rho^2)\ln(1-\rho^2)}{2\rho^2}-\ln\delta(\rho)\right],   
\end{equation}
where the vortex-core  cutoff $\delta(\rho)$ depends on the condensate density and is scaled with $R$. 

\begin{figure}[ht]
    \centering
    \includegraphics[width=0.8\linewidth]{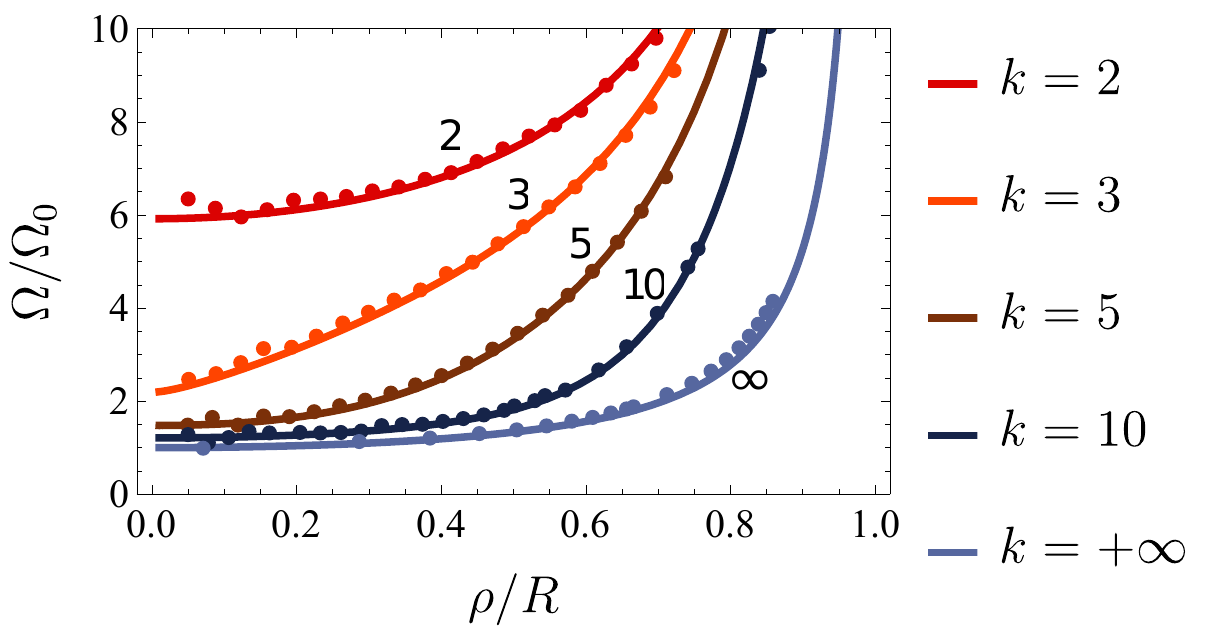}
    \caption{Dimensionless precession frequency $\Omega/\Omega_0$ [where $\Omega_0=\hbar/(m_aR^2)$] for a vortex located at distance $\rho$ from the center of a single-component BEC in a power-law $r^k$ trap. Solid lines are the analytic predictions of Eq.~\eqref{Omegaa}, while dots show results from the full time-dependent GP equation. 
    \rev{The numbers next to the lines denote the corresponding value of $k$.}
    } 
    \label{fig:Omegas_sweep_k}
\end{figure}

Figure \ref{fig:Omegas_sweep_k} shows the precession rate of a vortex in a single-component BEC for selected integer values of $k$ as a function of the vortex position $\rho$.  The solid curves show Eq.~\eqref{Omegaa} for the massive point vortex model studied here.  The dots show the results from the time-dependent Gross-Pitaevskii equation (solved using variants of algorithms we employed in Ref.~\cite{Richaud2021}).
The close overlap between lines and dots confirms the accuracy of the time-dependent variational Lagrangian formalism.

For comparison with the study of the dynamics of a massive point vortex, which will be discussed in  Sec.~\ref{sec:massive}, 
it is valuable to rewrite Eq.~(\ref{La}) in vector form
\begin{equation}\label{Lavec}
    L_a = q\frac{\tau(\rho)}{\rho}\,\dot{\bm \rho} \times\hat{\bm \rho }\cdot \hat{\bm z} -\epsilon(\rho).
\end{equation}
  The corresponding canonical momentum  
\begin{equation}\label{canmom}
    \bm p_a =\frac{\partial L_a}{\partial \dot{\bm \rho}}= q\frac{\tau(\rho)}{\rho} \hat{\bm \rho}\times\hat{\bm z} = -q\frac{\tau(\rho)}{\rho} \hat{\bm \phi}
\end{equation}
is in the azimuthal direction, as expected for uniform circular motion.  Note that the angular momentum is $\bm l = \bm \rho\times \bm p_a = -q\tau(\rho) \hat{\bm z}$, as found earlier.

The  dynamics for this massless vortex follows from $\dot{\bm p}_a =\partial L_a/\partial \bm \rho$, which gives 
\begin{equation}
    \dot{\bm p}_a = q\left(\frac{\tau(\rho)}{\rho^2}-\frac{\tau'(\rho)}{\rho}\right)\left(\dot{\bm \rho} \cdot\hat{\bm \phi}\right)\,\hat{\bm \rho}-\epsilon'(\rho)\hat{\bm \rho}.
\end{equation}
For uniform circular motion at fixed $\rho$, 
\rev{we note that $d\hat{\bm \phi}/dt = -\dot\phi \,\hat{\bm \rho}$ and a combination with  Eq.~(\ref{canmom}) gives  
\begin{equation}
    \dot{\bm p}_a = q \frac{\tau(\rho)}{\rho}\dot\phi\,\hat{\bm \rho} = q\frac{\tau(\rho)}{\rho^2}\left(\dot{\bm \rho}\cdot \hat{\bm \phi}\right) \hat{\bm \rho},
    \end{equation}
    } so that some terms cancel.  As a result, we find 
    \begin{equation}
        0 = \left[-2q(1-\rho^k)\left(\dot{\bm \rho}\times\hat{\bm z}\cdot\hat{\bm \rho}\right) -\epsilon'(\rho)\right]\hat{\bm \rho}.
    \end{equation}
    The last term is the ``force'' arising from the negative gradient of the energy $\epsilon(\rho)$.  In this picture, the vortex moves to ensure that the total force vanishes, which is precisely the ``Magnus''  effect. The resulting precession frequency reproduces the result given in Eq.~(\ref{Omegaa}), which follows more directly from the Lagrangian dynamics.

\section{Massive point vortex model}
\label{sec:massive}
\rev{In the presence of a second component-$b$, the  binary condensate obeys two coupled GP equations:
\begin{equation}
\label{eq:GPE_2_components}
i\hbar\frac{\partial \Psi_i}{\partial t} =\! \!\left(\!-\frac{\hbar^2\nabla^2}{2m_i} + V_{\rm tr} 
+ \sum_{j=a,b}g_{ij}|\Psi_j|^2\!\right)\Psi_i, \; i=a,b
\end{equation}
where $g_{ij}=\sqrt{2\pi}\hbar^2 a_{ij}/(m_{ij} d_z)$ represents the effective interactions in the quasi-2D system, with $a_{ij}$ being the intra- and inter-component $s$-wave scattering lengths and $m_{ij}=(m_i^{-1}+m_j^{-1})^{-1}$ are the reduced atomic masses \cite{Hadzibabic2011}. 
In the immiscible regime $g_{ab}>\sqrt{g_{aa}g_{bb}}$, the coupled equations (\ref{eq:GPE_2_components}) admit solutions where vortices are present in component-$a$ and wavepackets of $b$-particles are trapped within the vortices' cores. The dynamics of these composite objects (which we term ``massive vortices") can be conveniently described with an effective particle-like model \cite{Richaud2020,Richaud2021} which allows one to bypass the numerical solution of the GP equations (\ref{eq:GPE_2_components})}.

In our model for a massive point vortex, the total Lagrangian is the previous $L_a$ for the $a$ component augmented by the Lagrangian $L_b$ for the $b$ component. \rev{We remark that the presence of $N_b\ll N_a$ atoms within the vortex core does not significantly change its shape and width (see Ref.~\cite{Richaud2020} for the case of a $^{23}{\mathrm{Na}} - {}^{39}\mathrm{K}$ mixture and Ref.~\cite{Richaud2022_Collisions} for the case of a $^{87}{\mathrm{Rb}} - {}^{41}\mathrm{K}$ mixture), so that Eq.~(\ref{Psia}) remains valid.} As discussed in detail in Ref.~\cite{Richaud2021}, $L_b$ is proportional to $N_b$, with a kinetic term $\frac{1}{2}N_bm_b\dot{\bm \rho}^2$. The new feature here is the trap potential, so that $L_b = \frac{1}{2}N_bm_b\dot{\bm \rho}^2 -N_bV_{\rm tr}(\rho)$.  With our dimensionless variables, $L_b$ is
\begin{equation}
    L_b = \textstyle{\frac{1}{2}}{\mathfrak m}\dot{\bm \rho}^2 - \nu \rho^k  = \textstyle{\frac{1}{2}}{\mathfrak m}(\dot{ \rho}^2 + \rho^2\dot\phi^2)- \nu \rho^k,
\end{equation}
where ${\mathfrak m} =  M_b/M_a = N_bm_b/(N_am_a)$ is the ratio of the total $b$ mass to the total $a$ mass, and $\nu = N_b g_{aa} m_a/(\hbar^2\pi)$.  Note that $\nu$ is proportional to the product  $N_b g_{aa}$ and independent of $n_0$.

\rev{Our massive point-vortex model is expected to describe correctly the dynamics of a  single massive vortex  (and also a few massive vortices) as long as the component-$b$ atoms remain localized in the component-$a$ vortex cores, 
and as long as the $b$-atoms do not significantly alter the typical core size of component-$a$ bare vortices so that their cores remain much smaller than the TF radius $R$. The assumed immiscibility of the two components ensures the first condition, and the second one holds provided that $g_{aa}N_a$ is substantially larger than both $ g_{ab}N_b$ and $g_{bb}N_b$.
Note that the ratio $g_{bb}/g_{aa}$ does not enter the effective point-like model explicitly.}

\subsection{Total Lagrangian for massive point vortex}
\label{sub:Total_Lagrangian_for_massive_point_vortex}

In this way the Lagrangian $L$ for a single massive point vortex becomes
\begin{equation}\label{Ltotcoord}
    L = \textstyle{\frac{1}{2}}{\mathfrak m} (\dot{\rho}^2 + \rho^2\dot\phi^2) -q\tau(\rho)\dot\phi- \epsilon(\rho) - \nu \rho^k,
\end{equation}
here written in coordinate form.  It is helpful also to rewrite $L$ in vector form as  
\begin{equation}\label{Ltotvec}
    L = {\textstyle{\frac{1}{2}}}{\mathfrak m} \dot{\bm \rho}^2  +q\bm A(\rho)\cdot \dot{\bm \rho}- \epsilon(\rho) - \nu \rho^k,
\end{equation}
where 
\begin{equation}\label{A}
    \bm A(\bm \rho) = -\frac{\tau(\rho)}{\rho}\hat{\bm \phi}.
\end{equation}
\rev{The Lagrangian  has an unusual structure with a term linear in the  velocity $\dot{\bm \rho}$} in addition to the usual quadratic term proportional to the inertial mass.  

Such a Lagrangian is reminiscent of the Lagrangian $L_{\rm em}$ for a charged particle  at $\bm \rho$ in an external electromagnetic field
\begin{equation}\label{Lem}
    L_{\rm em} = {\textstyle{\frac{1}{2}}}{\mathfrak m}\dot{\bm \rho}^2 + q\dot{\bm \rho}\cdot \bm A -q\,\Phi,
\end{equation}
where $q$ is the charge, $\bm A$ is the vector potential and $\Phi$ is the scalar potential. 
The first two terms of Eqs.~(\ref{Ltotvec}) and (\ref{Lem}) are the same.  
 The third term of Eq.~(\ref{Ltotvec}) arises from the energy (\ref{Evor}) of the vortex and its image. It  is quadratic in the vortex charge and hence proportional to $q^2$, which here is simply $1$.  The last term in (\ref{Ltotvec}) is  the trap potential and hence independent of $q$.

 Equation (\ref{Ltotvec}) gives the canonical momentum 
 \begin{equation}\label{p}
     \bm p = \frac{\partial L}{\partial \dot{\bm \rho}} = {\mathfrak m}\dot{\bm \rho} +q\bm A.
 \end{equation}
 The corresponding Hamiltonian 
 \begin{equation}\label{H}
     H = \frac{\left(\bm p - q\bm A\right)^2}{2{\mathfrak m}} + \epsilon + \nu\rho^k
 \end{equation}
 is independent of  $t$, so that $H$ is constant (here, it is the energy, expressed in Hamiltonian variables $\bm p$  and $\bm \rho$).  
 
 Equation~(\ref{H}) identifies $\bm A$ in (\ref{A}) as a synthetic (or artificial) gauge field acting on the massive vortex.  Note that $\tau$  appearing in $\bm A$ involves an integral over the TF density $n_a(\rho)$. 
 Although we here study a power-law trap, other more general \rev{cylindrically symmetric TF densities could in principle be generated~\cite{Dalibard_DMD_2021}, leading to different forms for $\tau(\rho)$ and hence for $\bm A$.}  
  
The corresponding synthetic  magnetic field is
\begin{equation}
\bm B(\bm \rho) =\bm\nabla\times \bm A(\bm\rho) = -2\pi \hbar n_a(\rho)  \hat{\bm z},
\end{equation}
 here expressed in conventional units\footnote{\rev{To help understand the negative sign, consider a long solenoid with a uniform internal axial magnetic field along $\hat{\bm z}$, surrounding a uniformly charged dielectric cylindrical core with outward radial electric field.  The vector product $\bm E \times \bm B$ is along $-\hat{\bm \phi}$, and the subsequent vector product with $\bm r$ is along $-\hat{\bm z}$. }}.
 It is nonuniform except for a flat trap  ($k\to\infty$), but the total flux obtained as $\int d^2 \rho\, B_{\rm eff}(\bm \rho) = -2\pi \hbar N_a$ is independent of $k$.

\subsection{Dynamics of a massive point vortex}
\label{sub:Dynamics_of_a_massive_point_vortex}\rev{Equation~(\ref{p}) shows that the canonical momentum has an extra term proportional to the synthetic gauge field $\bm A$. An important consequence  is the presence of synthetic angular momentum, even for a massless vortex in a flat potential with a circular boundary, as noted in Sec.~II.A of~\cite{Richaud2021}.  In the present case of a massive point vortex, we now show that its angular momentum has the usual term proportional to the mass and the angular velocity, but it  also includes a second term arising from the synthetic gauge field.  Similar contributions are common in electromagnetism~\cite{PurcellBook, JacksonBook}.}
  
Since $L$ is independent of $\phi$, the angular momentum $l=\partial L/\partial \dot\phi$ of the massive vortex is conserved, with \begin{equation}\label{l}
  l =   {\mathfrak m} \rho^2\dot\phi -q\tau(\rho).
\end{equation}
Unlike the massless case, the angular momentum of a massive vortex can now have either sign.  
\rev{This unusual feature arises from the synthetic gauge field $\bm A = -[\tau(\rho)/\rho]\hat{\bm \phi}$. Here, the synthetic contribution is negative for a positive vortex, so that the total angular momentum can be negative  for a  vortex  precessing uniformly in the positive direction.  In addition, the angular momentum $l$ can vanish even for a precessing vortex, again owing to the synthetic contribution.}

The associated radial motion follows directly as 
\begin{equation}\label{rddot}
    {\mathfrak m} \ddot{\rho} = \frac{\partial L}{\partial \rho} = {\mathfrak m} \rho\dot{\phi^2} -q\tau'(\rho)\dot\phi -\epsilon'(\rho) -k\nu \rho^{k-1}.
\end{equation}
Some manipulation gives the energy equation
\begin{equation}\label{Econs}
 \frac{1}{2}{\mathfrak m} \dot \rho^2 + V_{\rm eff}(\rho) = \hbox{constant},
 \end{equation}
 with the effective potential 
\begin{equation}\label{Veff}
    V_{\rm eff}(\rho) = \frac{[l+q\tau(\rho)]^2}{2{\mathfrak m} \rho^2} + \epsilon(\rho)+ \nu \rho^k.
\end{equation}
Equation (\ref{Econs}) expresses the conservation of energy in Lagrangian variables $\dot{\bm\rho}$ and $\bm \rho$, which are generally more useful than the Hamiltonian variables. \rev{ In particular, the general radial dynamical equation becomes 
\begin{equation}\label{Newton}
    {\mathfrak m} \ddot{\rho} = - V_{\rm eff}'(\rho),
\end{equation}
balancing the Newtonian acceleration ${\mathfrak m}\ddot \rho$ and the force  $- V'_{\rm eff}(\rho)$.}

\begin{figure}[h!]
    \centering
    \includegraphics[width=0.8\linewidth]{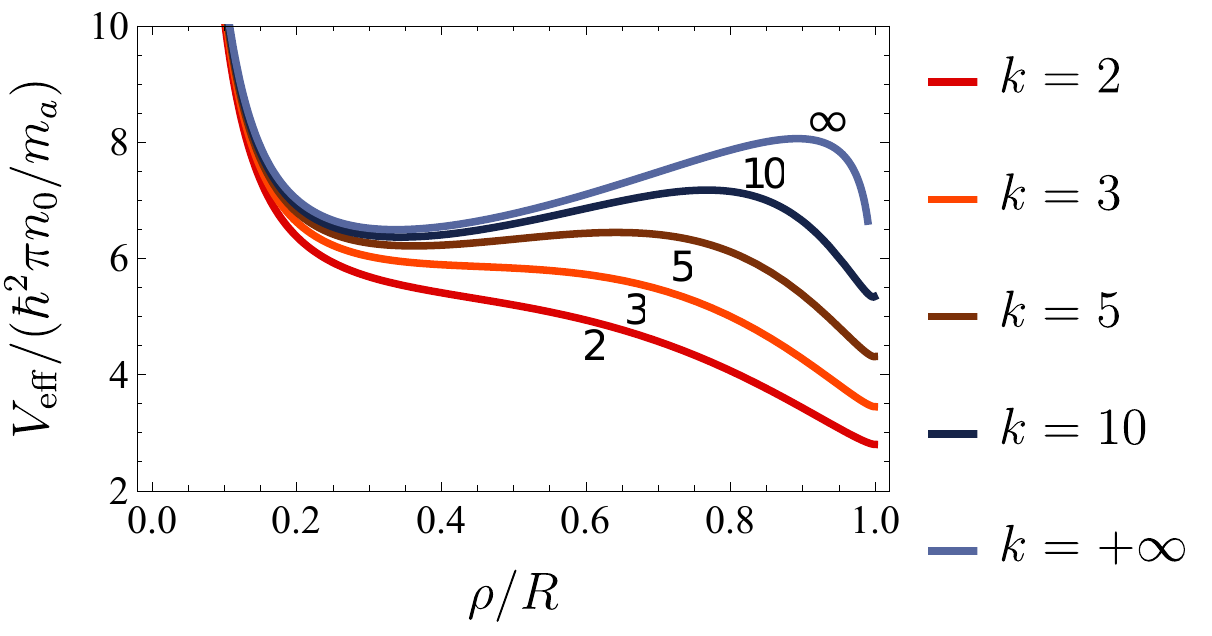}
    \caption{Plot of the effective potential (\ref{Veff}) as a function of the radial distance $\rho$, here shown for several values of $k$ and fixed $l=\mathfrak{m}=0.1$ and $\nu=1$. 
    \rev{The numbers next to the lines denote the corresponding value of $k$.}
    }
    \label{fig:Veff_vs_rho}
\end{figure}

\begin{figure}[h!]
    \centering
    \includegraphics[width=1\linewidth]{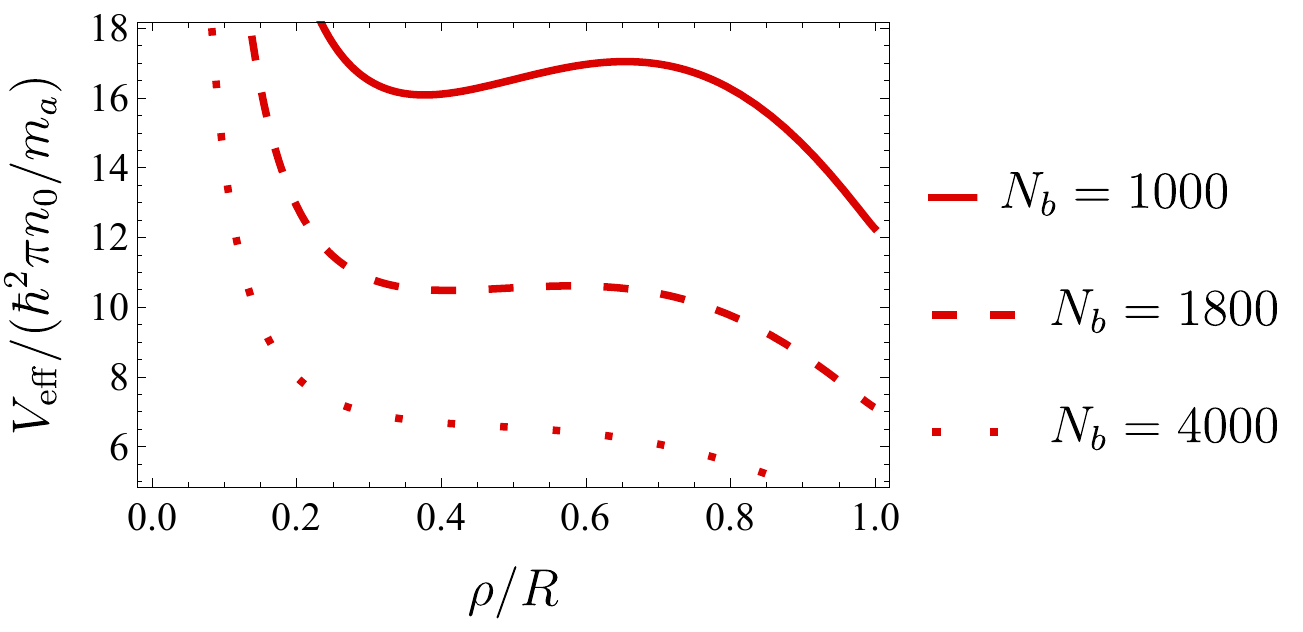}
    \caption{Plot of the effective potential (\ref{Veff}) as a function of the radial distance $\rho$, here expressed for several values of $N_b$ and fixed $k=2$, $l=0.1$,  $\mathfrak{m}=1.5\times 10^{-5} N_b$, $\nu=2.5\times 10^{-4} N_b$.}
    \label{fig:Veff_k_2}
\end{figure}

In many cases,  $V_{\rm eff}(\rho)$ has a single local minimum at a position {$\rho_0$} that depends on the  parameters $l$, ${\mathfrak m}$, and $\nu$.  Figure~\ref{fig:Veff_vs_rho} shows that the presence of  the minimum depends sensitively  on $k$. For a given  mass ratio $\mathfrak m$ the flat trap ($k\to\infty$) has a local minimum, but the latter disappears when $k$ decreases beyond a critical value.
The minimum also disappears as the number $N_b$  of atoms in the core increases, as shown in  Fig.~\ref{fig:Veff_k_2}.

 {For small $\mathfrak{m}\ll 1$, we can ignore the last two terms of (\ref{Veff}) and focus on the first term.  If  $l$ is also small, the minimum occurs at small $\rho_0 \approx \sqrt{l}$, confirming that there is always a local minimum, as expected from the behavior for a pure $a$ condensate.}
 
If the effective potential has a local minimum at $\rho_0$, 
a massive point vortex
at this radial distance from the origin precesses uniformly at a rate obtained by setting the right side of Eq.~(\ref{rddot}) to zero.  The resulting precession frequency $\Omega$ now satisfies a quadratic equation
\begin{equation}\label{Omegasq}
   {\mathfrak m} \rho_0 \Omega^2 - q\tau'(\rho_0) \Omega -\epsilon'(\rho_0) - k\nu \rho_0^{k-1} = 0.
\end{equation}
The first and last terms arise from the presence of the vortex mass, since 
both ${\mathfrak m}$ and $\nu$ are proportional to $N_b$.  In contrast, the second and third terms are just those studied in the previous section, including both the Magnus effect and the variational energy $\epsilon(\rho)$. Thus Eq.~(\ref{Omegasq}) includes all the physics inherent in our combined Lagrangian $L= L_a + L_b$.  

For small $N_b$, both ${\mathfrak m}$ and $\nu$  are small, and the first and last  terms in (\ref{Omegasq}) become negligible.  In this limit, the single root of  Eq.~(\ref{Omegasq}) reproduces Eq.~(\ref{Omegaa}) for a massless vortex.

{As $N_b$ increases}, however, Eq.~(\ref{Omegasq})   has  two finite solutions; a  massive point vortex at a given radial distance  $\rho_0$  then has two distinct modes with different precession frequencies. 
With positive $q$, the larger root  
 \begin{equation}\label{Omegap}
    \Omega_+ = \frac{\tau'(\rho_0) +\sqrt{\tau'(\rho_0)^2 + 4{\mathfrak m} \rho_0\left[\epsilon'(\rho_0) + k\nu \rho_0^{k-1}\right]}}{2{\mathfrak m} \rho_0}
\end{equation}
  diverges for small  mass ratio ${\mathfrak m}\ll 1$.
 
In contrast, the smaller root $\Omega_-$ is more physically significant because it remains finite for small $\mathfrak{m}$
\begin{equation}\label{Omegam}
    \Omega_- = \frac{2\left[-\epsilon'(\rho_0) -k\nu \rho_0^{k-1}\right]}{\tau'(\rho_0) +\sqrt{\tau'(\rho_0)^2 + 4{\mathfrak m} \rho_0\left[\epsilon'(\rho_0) + k\nu \rho_0^{k-1}\right]}}.
\end{equation} 
The denominator of Eq.~(\ref{Omegam}) is generally positive (as discussed below, it can  be complex, which implies instability). In contrast,  the numerator can have either sign, depending on the vortex position $\rho_0$ and the  dimensionless parameter $\nu$, which depends linearly on $N_b$. 
The quantity $-\epsilon'(\rho_0)$ is positive (see Fig.~\ref{fig:Epsilon_vs_rho}), but $-k\nu \rho_0^{k-1}$ is negative.  For small $N_b$ and $\nu$, the energy term with $-\epsilon'$ dominates and the positive massive vortex precesses  in the positive sense.  For larger $\nu$, however, the derivative of the trap potential dominates, and a positive vortex now precesses in the negative direction.

\begin{figure}[h!]
    \centering
    \includegraphics[width=0.8\linewidth]{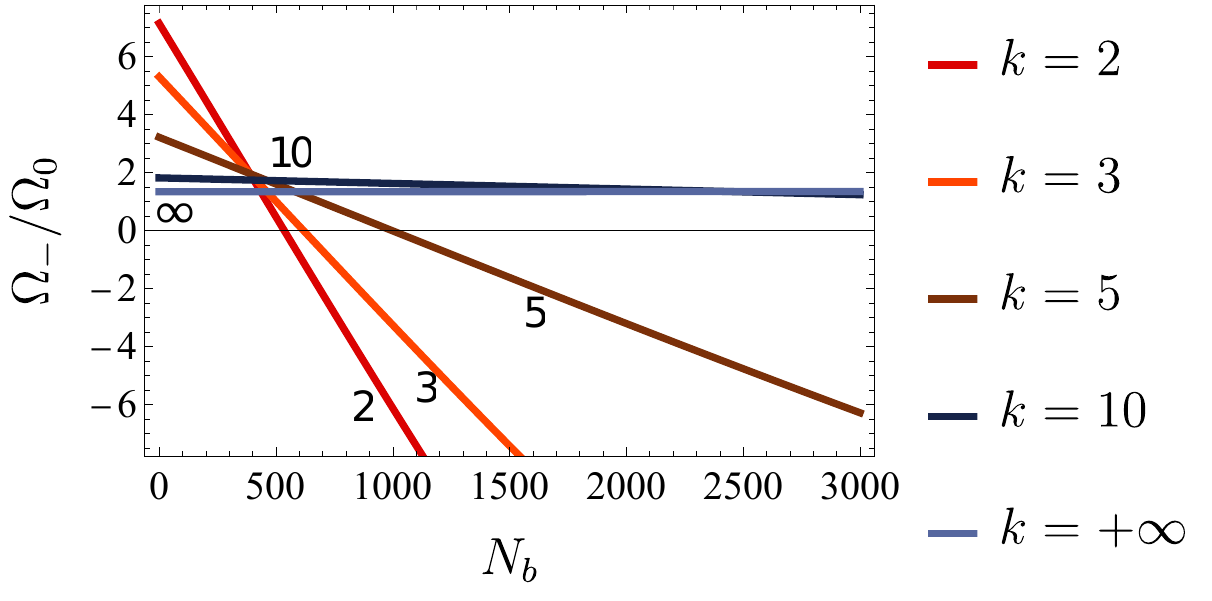}
    \caption{The precession frequency $\Omega_-$ [see Eq. (\ref{Omegam})] is positive for small core masses, but it can vanish and become negative as the number $N_b$ of component-$b$ core atoms increases. Here     $\mathfrak{m}=3\times 10^{-6}N_b$, $\nu=10^{-2}N_b$ and $\Omega_0=\hbar/(m_aR^2)$, which are typical values for a vortex in a BEC composed of $6\times 10^5$ $^{23}$Na atoms (pancake-shaped, with $R\sim 50~\mu$m and $d_z\sim 1~\mu$m), with the vortex mass provided by $N_b$ $^{39}$K atoms (see Ref. \cite{Zenesini_SciRep} for additional details about the tunability of intra- and interspecies coupling strengths in this specific Bose-Bose mixture). 
    \rev{The numbers next to the lines denote the corresponding value of $k$.}}
    \label{fig:Omega_minus_vs_Nb}
\end{figure}

Figure~\ref{fig:Omega_minus_vs_Nb} illustrates this situation for several  integer values of $k$.   Ruban \cite{Ruban2022} independently found similar behavior for $k = 2$  with a hydrodynamic model based on two  coupled GP equations.  In our model of a massive point vortex, the effect is  most pronounced for the harmonic trap ($k = 2$), and is absent for the flat trap ($k\to \infty$), as found in Ref.~\cite{Richaud2021}.

\rev{It is notable that the JILA two-component experiment with two hyperfine states of $^{87}$Rb (see Fig.~2 of~\cite{Anderson2000}) and the associated theoretical analysis (see Fig.~4 of ~\cite{McGeeHolland}) both found examples with negative precession frequencies, although it is not clear that they arose from the same mechanism.  In the experiment, the three interaction constants here are nearly equal, so that these results are definitely not in the regime of a well-localized core state.  It would be  desirable to have additional experiments  with two different atoms, such as our proposed $^{23}$Na and $^{39}$K mixture. }

\begin{figure}[h!]
    \centering
    \includegraphics[width=1\linewidth]{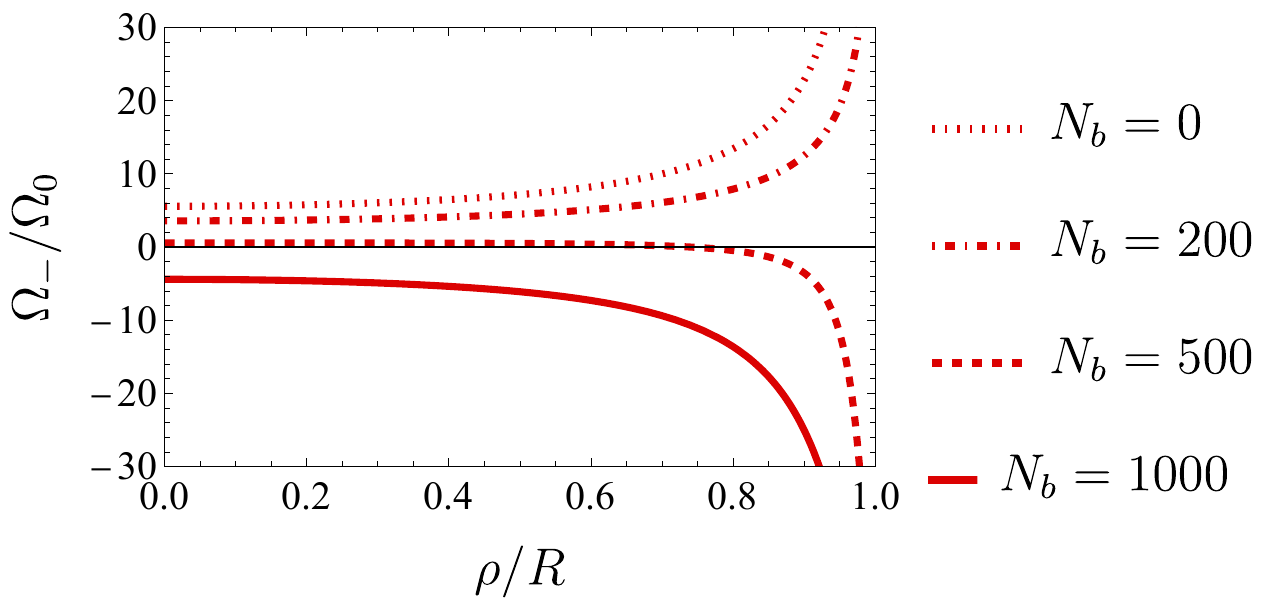}
    \caption{The precession frequency $\Omega_-$ [see Eq. (\ref{Omegam})] for $k=2$. It is positive for small core masses, but it can vanish and become negative as the number $N_b$ of component-$b$ core atoms increases. \rev{We employ the same microscopic model parameters used in Fig.~\ref{fig:Omega_minus_vs_Nb}, which yield $\mathfrak{m}=3\times 10^{-6} N_b$ and $\nu= 10^{-2} N_b$}. The normalization frequency $\Omega_0$ reads $\Omega_0=\hbar/(m_aR^2)$.}
    \label{fig:Omega_k_2_vs_r_sweep_Nb}
\end{figure}

\rev{The connection between the two roots $\Omega_\pm$ and the single local minimum $\rho_0$ may be understood by starting from 
parameters $\mathfrak{m}, l$ and $\nu$ that give}
a clear minimum such as the upper curve in Fig.~\ref{fig:Veff_k_2}. These values then yield $\Omega_\pm$ from Eqs.~(\ref{Omegap}) and (\ref{Omegam}). Use these frequencies to find the corresponding angular momenta $l_\pm$ from (\ref{l}). 
One of the solutions is the same as the input in finding the minimum of $V_{\rm eff}(\rho_0)$, but the other solution gives a second distinct $V_{\rm eff}(\rho_0)$. Although both effective potential curves have minima at the same $\rho_0$, they have different $l_\pm$ and therefore different detailed shapes.

Equation~(\ref{Omegasq}) has real coefficients, so that its roots are either real or complex conjugates, depending on the sign of the discriminant 
\begin{equation}\label{disc}
    {\cal D} = \tau'(\rho_0)^2 + 4{\mathfrak m} \rho_0\left[\epsilon'(\rho_0) + k\nu \rho_0^{k-1}\right].
\end{equation}
The quantity $\epsilon'(\rho_0)$ is negative 
and $k\nu \rho_0^{k-1} $ is positive, so that their sum can have either sign. 
The two roots $\Omega_\pm$ are usually real, but,  \rev{depending on the assumed values for the parameter $g_{aa}$ and the mass ratio $\mathfrak{m}$,} they can become complex-conjugate pairs, indicating that the vortex will not precess but instead drift to the outer boundary.  \rev{In Ref.~\cite{Richaud2021}  we found that  $\cal D$  in a flat trap was negative for $2{\mathfrak m} > 1-\rho_0^2$ ,} and we here generalize the discussion for general $k$.

\subsection{Stability of uniform precession for massive vortex}

If $V_{\rm eff}(\rho)$ has a local minimum at $\rho_0$, then the stability of the uniform precession follows by expanding around the minimum, with  $\rho = \rho_0 + \delta $.  Since  $V_{\rm eff}'(\rho_0)$ vanishes at a local minimum, the leading terms from Eq.~(\ref{Newton})  become
\begin{equation}
    {\mathfrak m} \ddot\delta + V_{\rm eff}''(\rho_0)\,\delta = 0.
\end{equation}
This equation describes a simple harmonic oscillator with squared frequency 
\begin{equation}
\label{eq:omega_oscillations}
 \omega^2 =\frac{V_{\rm eff}''(\rho_0)} {\mathfrak m},  
\end{equation}
where  the local curvature $V_{\rm eff}''(\rho_0)$ serves as an effective spring constant. 

\begin{figure}[h!]
    \centering
    \includegraphics[width=1\linewidth]{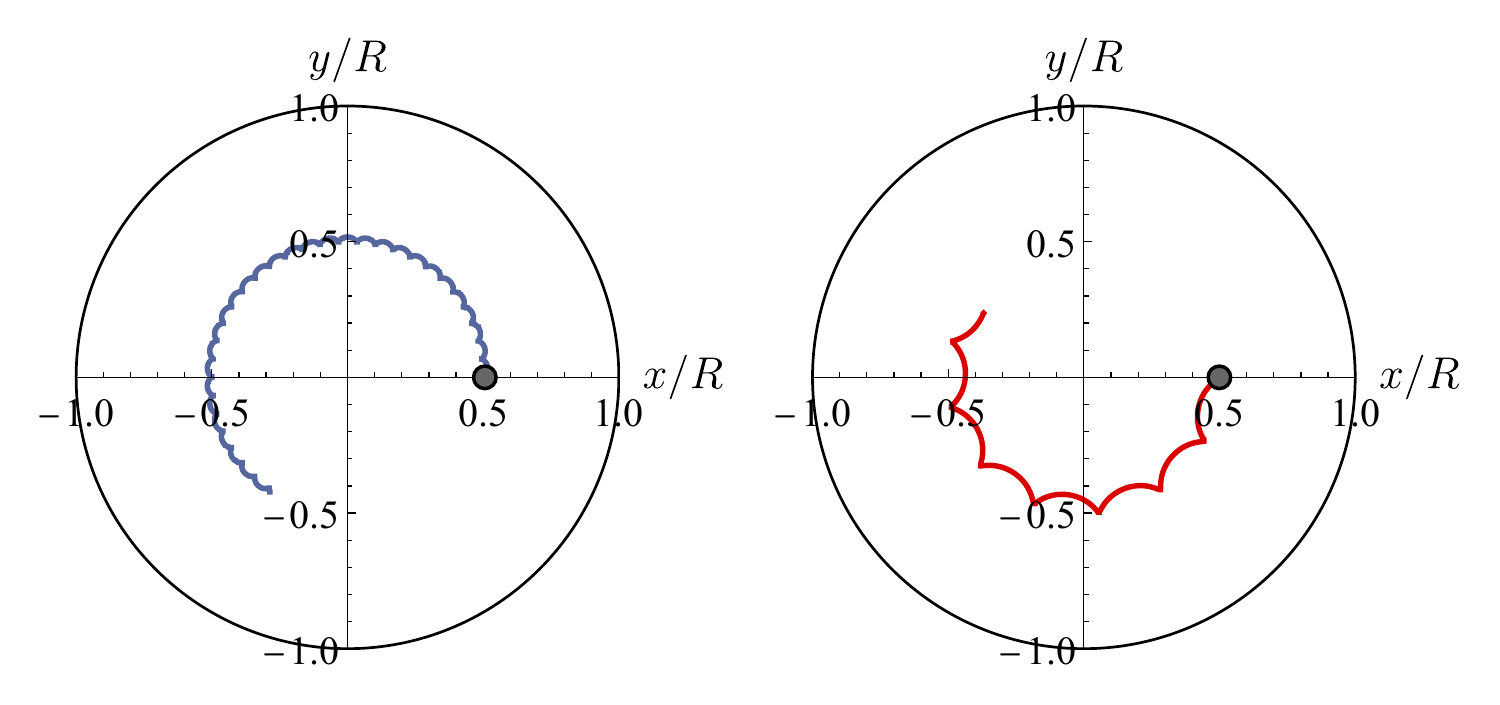}
    \caption{Perturbed uniform circular orbits of a positive vortex for  $k=\infty$ (left panel) and $k=2$ (right panel). Black dots correspond to the initial positions. The precession frequency, given by Eq.~\eqref{Omegam}, is positive in the first case and negative in the second.
    The frequency of small-amplitude radial oscillations is given by Eq.~(\ref{eq:omega_oscillations}). Here  we used $\mathfrak{m}=0.03$, $\nu=10$ and $\delta_0=0.005$.}
    \label{fig:Trajectories}
\end{figure}

An equivalent procedure is to expand the pair of Euler-Lagrange equations for $\rho$ and $\phi$ around the stable precessing motion, \rev{ with $\rho = \rho_0 +\delta\rho$ and $\phi = \Omega  t+ \delta\phi$}.  For example, the linearized form  of Eq.~(\ref{l}) gives 
\begin{equation}
    \left[2{\mathfrak m} \rho_0\Omega - q\tau'(\rho_0)\right] \delta \rho + {\mathfrak m} \rho_0^2 \delta\dot{\phi} = 0.
\end{equation}
With harmonic time dependence $\propto e^{-i\omega t}$, it is clear that the two perturbations $\delta\rho$ and $\delta \phi$ are out of phase because of the relative factor $i$.  A combination with the linearized form of the other dynamical equation (\ref{rddot}) readily gives the same oscillation frequency as in Eq.~(\ref{eq:omega_oscillations}).
Figure~\ref{fig:Trajectories} shows typical perturbed trajectories for both signs of the precession  frequency $\Omega_-$.

Figures~\ref{fig:Veff_vs_rho} and \ref{fig:Veff_k_2} indicate that the effective potential resembles a cubic curve with a single local minimum and a single local maximum.  The local minimum (maximum) is stable (unstable) with positive (negative) curvature.  As the parameters vary, the two stationary points can merge and form a single inflection point, which signals the onset of instability.  Beyond this point, the radial position obeys the simple Newtonian dynamical equation ${\mathfrak m} \ddot \rho  = -V_{\rm eff}'(\rho)$.
In this case, 
the vortex will move to the outer boundary of the condensate.  

\section{Conclusions and Outlook}
\label{sec:conclusions}

In this paper, we constructed a two-dimensional Lagrangian $L = L_a + L_b$ for a massive point vortex in a power-law trap potential $\propto r^k$. Our model assumes  a singly quantized vortex in condensate $a$  surrounding a localized condensate $b$ that provides an inertial mass. The power-law potential allows an interpolation between the  harmonic trap ($k = 2$) and the flat trap with a rigid circular boundary ($k \to\infty$). For an empty-core vortex in the pure $a$ component,  the  model Lagrangian $L_a$ leads to first-order dynamical equations with uniform circular precession for all values of $k$. 

To include the inertial effect of the core, we added the  Lagrangian $L_b$ derived in~\cite{Richaud2021}, now generalized to the power-law trap.  The total Lagrangian for the vortex coordinate $\bm \rho = (\rho,\phi)$ is axisymmetric and therefore conserves the total angular momentum $l$.  Unusually, $l$ includes not only the usual Newtonian inertial part $\propto \bm \rho\times \dot{\bm \rho} =  \rho^2\dot\phi$ but also a contribution from  a synthetic gauge field associated with the vortex.  
It is notable that synthetic gauge fields have served to create massless vortices \cite{Lin2009}, whereas here we identify a (density-dependent) synthetic gauge field that acts on a massive point vortex.  DMDs (digital micromirror devices)~\cite{Dalibard_DMD_2021} allow experiments with almost arbitrary condensate shapes, including time periodic structures. The corresponding time-periodic synthetic gauge fields could combine superfluid vortex dynamics with Floquet physics \cite{Eckardt2017} and Thouless pumping \cite{Ozawa2019}.

Manipulation of the coupled dynamical equations for $\rho$ and $\phi$ leads to an effective potential $V_{\rm eff}(\rho)$ and an explicit radial dynamical equation ${\mathfrak m} \ddot\rho = -V_{\rm eff}'(\rho)$, where ${\mathfrak m} = M_b/M_a$ is the mass ratio.  
For small enough values of $\mathfrak m$ and $l$, 
$V_{\rm eff}(\rho)$ has a single local minimum, where stable uniform circular motion can occur.  
For larger $\mathfrak m$, however, the local minimum disappears and the vortex spirals outward to the trap edge.

We studied the precession of a massive vortex for various values of the parameters in the Lagrangian such as the mass ratio ${\mathfrak m} =M_b/M_a$, the coupling strength  $\nu = N_b g_{aa} m_a/(\hbar^2\pi) $ between the vortex and the trap, and the exponent $k$ of the trap potential. 
For a flat potential ($k\rightarrow\infty$),
a positive massive vortex always precesses in the positive sense, independent of the number $N_b$ of $b$-component atoms which provide its mass.  
For finite $k$, however,  the precession can reverse direction with increasing $N_b$. The effect is stronger for smaller $k$ and therefore should be most easily observable in the usual case of harmonic trapping ($k = 2$). 


 \rev{As noted toward the end of Sec.~III.B, the early JILA experiment~\cite{Anderson2000} detected several two-component vortices that   precessed in the reverse direction.  These experiments relied on two hyperfine states of $^{87}$Rb, where the  interaction constants $g_{jk}$ are nearly identical.  In contrast, our model assumes different atomic species $^{23}$Na and $^{39}$K with the conditions $g_{aa}g_{bb}\ll g_{ab}^2$ to be deep in the phase-separated regime and $g_{aa}N_a\gg g_{ab}N_b,\,g_{bb}N_b$ to ensure that the size of a vortex in component $a$  is  barely modified by the $b$ impurities in its core. It would be very interesting to study vortices in  two-component  systems with small core radii and perhaps detect the reversal of precession as the minority component increases. }

\section*{Acknowledgements}
We thank Carlo Beenakker, Matteo Ferraretto, Giacomo Roati, Francesco Scazza, and Leticia Tarruell for stimulating discussions. P.M. was supported by grant PID2020-113565GB-C21 funded by MCIN/AEI/10.13039/501100011033, by the National Science Foundation under Grant No. NSF PHY-1748958, and by the {\it ICREA Academia} program.

\end{document}